\documentclass[two column, prA]{revtex4}%
\usepackage{amsmath}
\usepackage{graphicx}
\usepackage{amsfonts}
\usepackage{amssymb}
\usepackage{color}%
\providecommand{\U}[1]{\protect\rule{.1in}{.1in}}
\ifx\pdfoutput\relax\let\pdfoutput=\undefined\fi
\newcount\msipdfoutput
\ifx\pdfoutput\undefined\else
\ifcase\pdfoutput\else
\ifx\paperwidth\undefined\else
\ifdim\paperheight=0pt\relax\else\pdfpageheight\paperheight\fi
\ifdim\paperwidth=0pt\relax\else\pdfpagewidth\paperwidth\fi
\fi\fi\fi
\begin{document}
\title{Maxwell quantum mechanics}
\author{Margaret Hawton}
\affiliation{Department of Physics, Lakehead University, Thunder Bay, ON, Canada, P7B 5E1}
\email{margaret.hawton@lakeheadu.ca}

\begin{abstract}
We extend classical Maxwell field theory to a first quantized theory of the
photon by deriving a conserved Lorentz four-current whose zero component is a
positive definite number density. Fields are real and their positive
(negative) frequency parts are interpreted as absorption (emission) of a
positive energy photon. With invariant plane wave normalization, the photon
position operator is Hermitian with instantaneously localized eigenvectors
that transform as Lorentz four-vectors. Reality of the fields and wave
function ensure causal propagation and zero net absorption of energy in the
absence of charged matter. The photon probability amplitude is the real part
of the projection of the photon's state vector onto a basis of position
eigenvectors and its square implements the Born rule. Manifest covariance and
consistency with quantum field theory is maintained through use of the
electromagnetic four-potential and the Lorenz gauge.

\end{abstract}
\maketitle

\section{Introduction}

One can interpret the Dirac equation either as giving the dynamics for a
classical field or a quantum wave function \cite{Sebens}. The question
addressed here and in \cite{Sebens} is whether Maxwell's equations can be
interpreted as a first quantized theory of the photon. We answer this question
in the affirmative and show that classical electromagnetism in the Lorenz
gauge can be extended to a quantum mechanical theory that satisfies all of the
usual rules. A first quantized theory of a relativistic particle should have a
Lorentz covariant wave equation, a scalar lagrangian, future orientation and
causal propagation \cite{Kazemi,Kiessling}. In addition it should have a
conserved probability current that transforms like a Lorentz vector and
Hermitian operators representing observables that are not part of classical
field theory.

We waive the conditions of complex fields and a first order wave equation
stated in \cite{Kiessling} since, according to the Hegerfeldt theorem
\cite{Hegerfeldt}, this is incompatible with causal propagation. Hegerfeldt
proved that any positive frequency field localized for an instant spreads
instantaneously to give a field that is nonzero throughout space. In classical
electromagnetic theory fields are real and propagate causally, and in
relativistic quantum mechanics the fields describing \textit{neutral} bosons
are real \cite{ItzyksonZuber,MandlShaw}. Our solution to the causality problem
is to formulate photon quantum mechanics in terms of real fields. This
requires a number density that is positive definite for both positive and
negative frequency fields. Such an inner product does indeed exist, and its
properties were investigated by Mostafazadeh and co-workers
\cite{MostafazadehZamani,BabaeiMostafazadeh}. Here we extend their photon
density to a four-current that transforms as a Lorentz vector. We interpret
positive frequency as absorption of positive energy particles and negative
frequency as emission of positive energy particles; negative energy photons or
retrocausal propagation are not required.

An important contribution to the development of photon quantum mechanics is
investigation of the simpler Klein-Gordon equation describing neutral pions.
With the exception of the work of Mostafazadeh and co-workers, this has been
largely based on positive frequency solutions to a wave equation that is a
first order in time \cite{Kazemi,RosenHorwitz,Sutherland} with position
eigenvectors that are nonlocal and noncovariant
\cite{NewtonWigner,RosenHorwitz} or nonorthogonal \cite{Phillips,Sutherland}.
An exception is Wharton's formulation that imposes initial and final
conditions \cite{Wharton}. Work on the photon has followed a different but
sometimes interwoven path that was recently reviewed by Sebens \cite{Sebens}
and by Kiessling and Tarvildar-Zadeh \cite{Kiessling} in their Appendix A.
They propose a first order Dirac-like equation for a single free photon and
show that it satisfies all of the requirements outlined at the end of
paragraph one.

Up to the turn of this century there was no photon number density and no true
photon quantum mechanics. This is a consequence of several obstacles to its
development: Newton and Wigner \cite{NewtonWigner} derived position operators
for electrons and for spin 0 Klein-Gordon particles but found that, if
invariance under rotations is assumed, no photon position operator exists.
Even these Klein-Gordon position eigenvectors are problematic since they are
not Lorentz covariant and they are nonlocal in configuration space. Pryce
\cite{Pryce} derived a photon position operator but its components do not
commute, so it does not define a basis of position eigenvectors. The standard
density is positive only if the fields are restricted to positive frequencies
however, as discussed in a previous paragraph, positive frequency fields alone
do not propagate causally \cite{Hegerfeldt}. Some form of a photon wave
function is thought to be useful in quantum optics but, as a consequence of
the above mentioned obstacles, the usual rules of quantum mechanics were not
followed. Instead the Weber vector \cite{Weber,Kiessling} discussed by
Bialynicki-Birula \cite{BB96} and the Sipe photon wave function \cite{Sipe}
are based on energy density \cite{SmithRaymer}. Alternatively the Weber vector
can be divided by the square root of frequency to give number amplitude
\cite{Good}. This wave function was examined recently by Sebens \cite{Sebens},
but found to be unsatisfactory because probabilities do not always transform
properly under Lorentz transformations.

The question of localizability is fundamental to quantum field theory (QFT)
where particles interact locally at a common space-time coordinate. Particle
position, while essential to this picture, is especially problematic in the
case of photons that have no rest frame. Recently, derivation of a photon
position operator with commuting components \cite{HawtonPosOp,HawtonBaylis}
and the Mostafazadeh particle density that is positive definite for both
positive and negative frequency waves \cite{MostafazadehZamani} have led to
photon quantum mechanics \cite{HawtonDebierre,BabaeiMostafazadeh} with a
Hilbert space consisting of solutions to Maxwell's equations and an invariant
inner product. However, this is not an entirely satisfactory first quantized
theory since in \cite{BabaeiMostafazadeh} the position operator and its
eigenvectors are of the noncovariant Newton-Wigner form, while in
\cite{HawtonDebierre} the position operator is not Hermitian and
biorthogonality is derived from QFT. Also, in
\cite{BabaeiMostafazadeh,HawtonDebierre} and in \cite{Sebens}, no photon
density that transforms like a Lorentz vector was found. Here we derive a
photon number density that transforms as the zeroth component of a Lorentz
vector. In addition, we show that invariant plane wave normalization leads to
a Hermitian photon position operator with localized eigenvectors that
transform as Lorentz vectors.

In the next Section {we give} a brief summary of relativistic notation and
some material from the QFT text book \cite{ItzyksonZuber} that motivated our
work. {Since we discuss only photons and Fermions, quantum electrodynamics
(QED) as discussed in QFT text books is sufficient for our purposes.} In
Section III we write the general solution to Maxwell's wave equation as a sum
over positive and negative frequency terms ($\epsilon=\pm$) in the Lorenz
gauge, define the Babaei-Mostafazadeh conjugate field, and derive a photon
four current. An invariant inner product is then defined in cofiguration space
and in momentum space. In Section III.B the photon position operator with
commuting components is reviewed. It is found that its eigenvectors are
localized Lorentz vectors. An expression for probability amplitude is obtained
by projecting onto a basis of position eigenvectors and taking its real part.
SI units are used throughout.

\section{Background: Relativity and QFT}

\subsection{Relativistic notation}

In relativity the space-time coordinates are incorporated into the 4-vector
$x=x^{\mu}=\left(  ct,\mathbf{x}\right)  $ with Lorentz invariant length
squared $x^{2}=x_{\mu}x^{\mu}$ where $x_{\mu}=g_{\mu\nu}x^{\nu}$ and
$g_{\mu\nu}=g^{\mu\nu}$ is a $4\times4$ matrix with diagonal $\left(
1,-1,-1,-1\right)  $. The classical (in the non-quantum sense) energy-momemtum
four-vector is $p^{\mu}=\left(  E_{p}/c,\mathbf{p}\right)  =\hbar\left(
\omega_{k}/c,\mathbf{k}\right)  $ where we define $\omega_{k}$ to be positive.
The four-potential is $A^{\mu}=\left(  \phi/c,\mathbf{A}\right)  $ where
$\phi$ and $\mathbf{A}$ are the scalar and vector potential respectively. In
first quantization energy and momentum are replaced with operators, that is
$E_{p}\rightarrow$\textrm{i}$\hbar\partial_{t}$ and $\mathbf{p}\rightarrow
\widehat{\mathbf{p}}=-$\textrm{i}$\hbar\partial_{\mathbf{x}}$. Here $E_{p}$ is
the energy of a particle with momentum $\mathbf{p},$ $c$ is the speed of
light, $h=2\pi\hbar$ is Planck's constant, $\partial_{t}\equiv\partial
/\partial t$ and $\partial_{\mathbf{x}}\equiv\nabla$. Some additional
invariants that will be used here are $p_{\mu}p^{\mu}=E_{p}^{2}/c^{2}%
-\mathbf{p}^{2}c^{2}$,
\begin{equation}
\square\equiv\partial_{\mu}\partial^{\mu}=\partial_{ct}^{2}-\boldsymbol{\nabla
}^{2}, \label{box}%
\end{equation}
$kx=\omega_{k}t-\mathbf{k\cdot x}$ and $\exp\left(  -\mathrm{i}kx\right)  $.
The $\mathbf{k}$-space Lorenz gauge condition, $kA=0$, is Lorentz invariant
while the Coulomb gauge condition, $\mathbf{k}\cdot\mathbf{A}=0$, is not so,
in keeping with our theme of maintaining manifest covariance, we will use the
Lorenz gauge. Operation with $\widehat{E}_{p}^{2}-\widehat{\mathbf{p}}%
^{2}c^{2}$ on the four-potential $A\left(  x\right)  =A\left(  t,\mathbf{x}%
\right)  $ gives the Maxwell wave equation,$\ $
\begin{equation}
\square A^{\nu}=0. \label{MaxwellWaveEq}%
\end{equation}
Since this equation is second order in both space and time and $m=0$, it has
the remarkable property that it is unchanged by multiplication by $\hbar^{2}$
so it can be interpreted as a classical or a quantum wave equation.

\subsection{Lessons from QFT text books}

Because QFT is based on local interactions it is common in QFT to use the
Lorenz gauge and the Gupta-Bleuler indefinite metric
\cite{GuptaBleuler,MandlShaw,ItzyksonZuber,HawtonWaveFunctions99}. This
approach is covariant and based on four essentially identical copies of the
solution to the $m=0$ Klein-Gordon wave equation $\square\phi=0$. The
Lagrangian is the usual one plus a term proportional to the square of the
Lorenz gauge condition, $\mathcal{L=-}\frac{1}{4\mu_{0}}F_{\mu\nu}F^{\mu\nu
}-\frac{1}{2\mu_{0}}\left(  \partial_{\mu}A^{\mu}\right)  ^{2}$ where $\mu
_{0}$ is the magnetic permeability of vacuum and $F^{\mu\nu}$ is the
electromagnetic tensor whose first row is $F^{0\nu}=\left(  0,\mathbf{E}%
\right)  /c$, $\mathbf{E}$ being the electric field vector. The field
conjugate to $A_{\nu}$ is $\partial\mathcal{L}/\partial\left(  \partial
_{ct}A_{\nu}\right)  =F^{0\nu}-g^{0\nu}\left(  \partial_{\mu}A^{\mu}\right)
=\pi^{\nu}$ with
\begin{equation}
\pi^{\nu}=-\epsilon_{0}\partial_{t}A^{\nu}. \label{photonpi}%
\end{equation}
In a source free region the Lagrange equation of motion is just Maxwell's wave
equation, (\ref{MaxwellWaveEq}).

The second quantized four-potential operator can be written as
\cite{ItzyksonZuber}%
\begin{align}
\widehat{A}^{\mu}\left(  x\right)   &  =\sum_{\lambda=0,3}\widehat{A}%
_{\lambda}^{\mu}\left(  x\right)  ,\nonumber\\
\widehat{A}_{\lambda}^{\mu}\left(  x\right)   &  =\sqrt{\frac{\hbar}%
{\epsilon_{0}}}\int\frac{\mathrm{d}\mathbf{k}}{\left(  2\pi\right)
^{3}2\omega_{k}}\left[  \widehat{a}_{\lambda}\left(  \mathbf{k}\right)
e_{\lambda}^{\mu}e^{-\mathrm{i}kx}\right. \nonumber\\
&  \left.  +\widehat{a}_{\lambda}^{\dag}\left(  \mathbf{k}\right)  e_{\lambda
}^{\mu\ast}e^{\mathrm{i}kx}\right]  \label{4potentialoperator}%
\end{align}
where $\widehat{a}_{\lambda}\left(  \mathbf{k}\right)  $ and $\widehat{a}%
_{\lambda}^{\dagger}\left(  \mathbf{k}\right)  $ are annihilation and creation
operators for a photon with wave vector $\mathbf{k}$ and polarization
$\lambda$ and $e_{\lambda}^{\mu}\left(  \mathbf{k}\right)  $ is a unit vector.
The first term in (\ref{4potentialoperator}) annihilates a photon at $x$,
while the second term creates one. The conjugate field operator is%
\begin{equation}
\widehat{\pi}_{\lambda}^{\mu}\left(  x\right)  =-\epsilon_{0}\partial
_{t}\widehat{A}_{\lambda}^{\mu}\left(  x\right)  \label{fieldoperator}%
\end{equation}

With the polarization unit vectors defined such that 0 is time-like, 1 and 2
are transverse and 3 is longitudual,%
\begin{equation}
e_{0}=n^{\mu}=\left(  1,0,0,0\right)  ,\ \mathbf{e}_{3}\left(  \mathbf{k}%
\right)  =\mathbf{e}_{\mathbf{k}}=\mathbf{k}/\left\vert \mathbf{k}\right\vert
, \label{e0}%
\end{equation}%
\begin{equation}
\mathbf{e}_{\lambda}^{\ast}\left(  \mathbf{k}\right)  \cdot\mathbf{e}%
_{\lambda^{\prime}}\left(  \mathbf{k}\right)  =\delta_{\lambda,\lambda
^{\prime}}\text{ for}\ \lambda,\lambda^{\prime}=1,2,3. \label{evector}%
\end{equation}
Defining%
\begin{equation}
\zeta=\left(  -1,1,1,1\right)  \label{squiggle}%
\end{equation}
this can be written as%
\begin{equation}
e_{\lambda\mu}^{\ast}e_{\lambda^{\prime}}^{\mu}=e_{\lambda0}^{\ast}%
e_{\lambda^{\prime}}^{0}-\mathbf{e}_{\lambda}^{\ast}\cdot\mathbf{e}%
_{\lambda^{\prime}}=-\zeta_{\lambda}\delta_{\lambda,\lambda^{\prime}}.
\label{es}%
\end{equation}
The operator version of the Lorenz gauge condition is then
\begin{equation}
\left[  \widehat{a}^{0}\left(  \mathbf{k}\right)  -\widehat{a}^{3}\left(
\mathbf{k}\right)  \right]  \left\vert 0\right\rangle =0. \label{Lorenz}%
\end{equation}
In the absence of an electric four-current density the number of longitudinal
photons must equal the number of scalar photons. The creation and annihilation
operators for a photon with wave vector $\mathbf{k}$ and helicity $\lambda$
will be assumed to satisfy the invariant commutation relations
\begin{align}
\left[  \widehat{a}_{\lambda}\left(  \mathbf{k}\right)  ,\widehat{a}%
_{\lambda^{\prime}}\left(  \mathbf{q}\right)  \right]   &  =0,\ \left[
\widehat{a}_{\lambda}^{\dagger}\left(  \mathbf{k}\right)  ,\widehat{a}%
_{\lambda^{\prime}}^{\dagger}\left(  \mathbf{q}\right)  \right]
=0,\nonumber\\
\left[  \widehat{a}_{\lambda}\left(  \mathbf{k}\right)  ,\widehat{a}%
_{\lambda^{\prime}}^{\dagger}\left(  \mathbf{q}\right)  \right]   &
=\delta_{\lambda,\lambda^{\prime}}\left(  2\pi\right)  ^{3}2\omega_{k}%
\delta\left(  \mathbf{k}-\mathbf{q}\right)  . \label{kcommutation}%
\end{align}
so that the one photon states%
\begin{equation}
\left\vert 1_{\mathbf{k}\lambda}\right\rangle =\widehat{a}_{\lambda}^{\dag
}\left(  \mathbf{k}\right)  \left\vert 0\right\rangle \label{1kphoton}%
\end{equation}
are invariantly normalized according to%
\begin{equation}
\left\langle 1_{\mathbf{k}\lambda}|1_{\mathbf{q}\lambda^{\prime}}\right\rangle
=\delta_{\lambda,\lambda^{\prime}}\left(  2\pi\right)  ^{3}2\omega_{k}%
\delta\left(  \mathbf{k}-\mathbf{q}\right)  . \label{korthogonal}%
\end{equation}

An annihilation operator acting on the vacuum state gives zero, that is
$\widehat{a}_{\lambda}\left(  \mathbf{k}\right)  \left\vert 0\right\rangle =0$
and $\widehat{A}_{\lambda}^{\mu+}\left(  x\right)  \left\vert 0\right\rangle
=0$. It can be proved using (\ref{4potentialoperator}), (\ref{fieldoperator}),
(\ref{es}), and (\ref{kcommutation}) that the $\mathbf{x}$-space photon
commutation relations at time $t$ are
\begin{align}
\left[  \widehat{A}_{\lambda}^{\mu}\left(  t,\mathbf{x}\right)  ,\widehat{A}%
_{\lambda^{\prime}\mu}\left(  t,\mathbf{y}\right)  \right]   &  =0,\ \left[
\widehat{\pi}_{\lambda}^{\mu}\left(  t,\mathbf{x}\right)  ,\widehat{\pi
}_{\lambda^{\prime}\mu}\left(  t,\mathbf{y}\right)  \right]  =0,\nonumber\\
\left[  \widehat{A}_{\lambda}^{\mu}\left(  t,\mathbf{x}\right)  ,\widehat{\pi
}_{\lambda^{\prime\prime}\mu}\left(  t,\mathbf{y}\right)  \right]   &
=-\mathrm{i}\hbar\zeta_{\lambda}\delta_{\lambda,\lambda^{\prime}}\delta\left(
\mathbf{x}-\mathbf{y}\right)  \label{APicommutation}%
\end{align}
where summation over $\mu$ is implied. The factor $\zeta_{\lambda}$ comes from
(\ref{es}). It was omitted in the definition (\ref{kcommutation}) to avoid
introducing it twice. A creation operator at $x$ acting on the vacuum state
gives the 1-particle state
\begin{equation}
\left\vert A_{x\lambda}^{\mu}\right\rangle =\sqrt{\frac{\hbar}{\epsilon_{0}}%
}\int\frac{\mathrm{d}\mathbf{k}}{\left(  2\pi\right)  ^{3}2\omega_{k}%
}e^{-\mathrm{i}kx}e_{\lambda}^{\mu}\left(  \mathbf{k}\right)  \left\vert
1_{\mathbf{k}\lambda}\right\rangle . \label{1photonatx}%
\end{equation}
The vacuum expectation value of (\ref{APicommutation}) can be written as
\begin{equation}
\left\langle A_{t,\mathbf{x}}^{\mu}|\pi_{t,\mathbf{y}\mu}\right\rangle
-\left\langle \pi_{t,\mathbf{y}\mu}|A_{t,\mathbf{y}}^{\mu}\right\rangle
=-\mathrm{i}\hbar\zeta_{\lambda}\delta_{\lambda,\lambda^{\prime}}\delta\left(
\mathbf{x}-\mathbf{y}\right)  . \label{VacuumExpectation}%
\end{equation}
with
\begin{equation}
\left\vert \pi_{x\lambda\mu}\right\rangle \equiv\widehat{\pi}_{\lambda\mu
}\left(  x\right)  \left\vert 0\right\rangle . \label{Pioperator}%
\end{equation}
Creation at $\left(  t,\mathbf{y}\right)  $ followed by annihilation at
$\left(  t,\mathbf{x}\right)  $ and creation at $\left(  t,\mathbf{x}\right)
$ followed by annihilation at $\left(  t,\mathbf{y}\right)  $ are equally
likely. This implies that we don't know whether the photon was created or
destroyed at $\left(  t,\mathbf{x}\right)  $.

We will illustrate Hegerfeldt instantaneous spreading by evaluating the
positive frequency part of (\ref{VacuumExpectation}). Integration of
$\left\langle A_{t,\mathbf{x}}^{\mu}|\pi_{\mu t,\mathbf{y}}\right\rangle $
after substitution of (\ref{1photonatx}), (\ref{Pioperator}) and
(\ref{korthogonal}) with $\mathbf{r\equiv x}-\mathbf{y}$ and $t\equiv
t_{x}-t_{y}$ gives \cite{HawtonDebierre}%
\begin{align}
\left\langle A_{t,\mathbf{x}}^{\mu}|\pi_{t,\mathbf{y}\mu}\right\rangle  &
=-\mathrm{i}\hbar\int\frac{\mathrm{d}\mathbf{k}e^{-\mathrm{i}\omega_{k}%
t}e^{\mathrm{i}\mathbf{k}\cdot\mathbf{r}}}{\left(  2\pi\right)  ^{3}%
2}\nonumber\\
&  =-\mathrm{i}\hbar\frac{1}{8\pi^{2}r}\frac{\partial}{\partial r}\sum
_{\gamma=\pm}\nonumber\\
&  \times\left[  \pi\delta\left(  r-\gamma\epsilon ct\right)  +\mathrm{i}%
\gamma PV\left(  \frac{1}{r-\gamma\epsilon ct}\right)  \right]  .
\label{deltaPV}%
\end{align}
At $t=0$ the principal value ($PV$) terms cancel while the $\delta$-functions
add to give a factor $2$. The $1/r$ tails of the $PV$ terms are
instantaneously masked by destructive interference \cite{Karpov}, only their
sum is localized. Eq. (\ref{deltaPV}) was included to illustrate the
nonlocality of the $PV$ terms but, since $\left\langle \pi_{t,\mathbf{y}\mu
}|A_{t,\mathbf{y}}^{\mu}\right\rangle =-\left\langle A_{t,\mathbf{x}}^{\mu
}|\pi_{t,\mathbf{y}\mu}\right\rangle ^{\ast}$, (\ref{VacuumExpectation}) that
is a sum of positive and negative frequency terms can be written as
\begin{equation}
\left\langle A_{t,\mathbf{x}}^{\mu}|\partial_{ct}A_{t,\mathbf{y}\mu
}\right\rangle -\left\langle \partial_{ct}A_{t,\mathbf{y}\mu}|A_{t,\mathbf{x}%
}^{\mu}\right\rangle =\mathrm{i}\frac{\hbar}{2\epsilon_{0}c}\zeta_{\lambda
}\delta_{\lambda,\lambda^{\prime}}\delta\left(  \mathbf{x}-\mathbf{y}\right)
. \label{QEDproduct}%
\end{equation}

To sum up, the photon equation of motion in the Lorenz gauge can be derived
from a scalar Lagrangian. Time-like, longitudinal, and transverse unit vectors
were defined. Derivation of the configuration space commutation relations
(\ref{APicommutation}) illustrates the role of invariant normalization of the
plane wave basis. The QFT commutation relations that ensure microcausality
require a sum of a term describing propagation from $y$ to $x$ and a term
representing propagation from $x$ to $y$, giving a real result. This implies
that emission at $y$ followed by absorption at $x$ and emission at $x$
followed by absorption at $y$ are equally likely; we don't know if a the
photon was emitted or absorbed at $x$.

\section{Maxwell quantum mechanics}

\subsection{Hilbert space}

A first quantized theory requires a vector space, a conserved four-current, an
inner product, a positive definite norm, and operators representing
observables. Here the photon vector space consists of solutions to the Maxwell
wave equation (\ref{MaxwellWaveEq}) subject to the Lorenz gauge condition
$\partial_{\mu}A^{\mu}=0$. In $\mathbf{k}$-space the Lorenz gauge condition
is
\begin{equation}
\left\vert \mathbf{k}\right\vert A_{\parallel}=\left(  \omega_{k}/c\right)
A^{0} \label{1stQuantizedLorenz}%
\end{equation}
so, since $\omega_{k}=c\left\vert \mathbf{k}\right\vert $ in vacuum, the
probability amplitude for a longitudinal mode equals the probability amplitude
for a scalar mode. Since (\ref{MaxwellWaveEq}) is second order in time,
completeness requires both positive and negative frequency solutions, that is
$A^{\mu}\left(  t,\mathbf{x}\right)  =\sum_{\lambda,\epsilon=\pm}A_{\lambda
}^{\epsilon\mu}\left(  t,\mathbf{x}\right)  $ where
\begin{equation}
A_{\lambda}^{\epsilon\mu}\left(  t,\mathbf{x}\right)  =\mathrm{i}\sqrt
{\frac{\hbar}{\epsilon_{0}}}\int\frac{\mathrm{d}\mathbf{k\ }}{\left(
2\pi\right)  ^{3}2\omega_{k}}c_{\lambda}^{\epsilon}\left(  \mathbf{k}\right)
e_{\lambda}^{\mu}\left(  \mathbf{k}\right)  e^{-{i}\epsilon\left(  \omega
_{k}-\mathbf{k}\cdot\mathbf{x}\right)  }. \label{Afield}%
\end{equation}
The unit imaginary \textrm{i }was introduced for later convenience to give the
integrand $\sin\left(  kx+\arg c^{+}\right)  $ that is an odd function of $kx$
if $c_{\lambda}^{\epsilon}\left(  \mathbf{k}\right)  $ is real. The
$\epsilon=+$ terms $e^{-\mathrm{i}\omega_{k}t}$ are referred to as positive
frequency while the $\epsilon=-$ terms $e^{\mathrm{i}\omega_{k}t}$ are called
negative frequency. Our sign convention is that positive $\epsilon$ signifies
absorption and negative $\epsilon$ denotes emission which can be thought of as
negative or time reversed absorption. Eq. (\ref{1photonatx}) is the state
vector describing a positive frequency one photon state in this vector space,
but there are no particle annihilation and creation operators in a first
quantized theory, so it cannot be attributed to the action of a creation
operator on the vacuum.

Babaei and Mostafazadeh \cite{BabaeiMostafazadeh} define the conjugate
field\textbf{\ }%
\begin{equation}
A_{c}\equiv\mathrm{i}\widehat{D}^{-1/2}\partial_{ct}A \label{Ac}%
\end{equation}
where%
\begin{equation}
{\widehat{D}\equiv-\nabla^{2}.} \label{D}%
\end{equation}
It can be verified by substitution that if $A$ satisfies Maxwell's wave
equation, $A_{c}$ also satisfies this equation.{ The operator on the right
hand side of (\ref{Ac}) extracts the sign of the photon frequency, }$\epsilon
$, according to%
\begin{equation}
\mathrm{i}\widehat{D}^{-1/2}\partial_{ct}A^{\epsilon}=\epsilon A^{\epsilon}.
\label{epsilon}%
\end{equation}

The inner product on a $t$-hyperplane should equal the spatial integral of the
$0^{th}$ component of a four-current, so the photon continuity equation will
be considered next. For any $A$ and $\widetilde{A}$ that satisfy
(\ref{MaxwellWaveEq}), a continuity equation can be obtained by subtracting
$\widetilde{A}$ multiplied by the Maxwell wave equation for $A^{\ast}$ from
$A^{\ast}$ times the Maxwell wave equation for $\widetilde{A}.$ This is the
method used to derive the Schr\"{o}dinger continuity equation in undergraduate
quantum mechanics. After cancellation of the $\partial_{\mu}A_{\nu}^{\ast
}\partial^{\mu}\widetilde{A}^{\nu}$ terms, substitution of
(\ref{MaxwellWaveEq}) and multiplication by the constant \textrm{i}$g$,
\begin{equation}
\mathrm{i}g\partial_{\mu}A_{\nu}^{\ast}\left(  x\right)
\overleftrightarrow{\partial}^{\mu}\widetilde{A}^{\nu}\left(  x\right)  =0
\label{Maxwellcontinuity}%
\end{equation}
where
\begin{equation}
f\overleftrightarrow{\partial}_{\mu}g\equiv f\left(  \partial_{\mu}g\right)
-g\left(  \partial_{\mu}f\right)  . \label{fg}%
\end{equation}
Eq. (\ref{Maxwellcontinuity}) is of the form
\begin{equation}
\partial_{t}\rho+\partial_{\mathbf{x}}\cdot\mathbf{j}=\partial_{\mu}J^{\mu}=0
\label{Continuity}%
\end{equation}
where $\rho$ is density, $\mathbf{j}$ is current density and $J^{\mu}$ is
four-current density. If $\widetilde{A}=\phi$ and $g=$ $e\epsilon_{0}c/\hbar$
where $e$ is the magnitude of the charge on the electron, $\rho$ is pion
charge density. While this charge density is not directly relevant to the
photon in vacuum, it provides a simple argument for selecting $g=\epsilon
_{0}c/\hbar$ to describe number density. If $\widetilde{A}=A_{c}$
(\ref{Maxwellcontinuity}) is a continuity equation for four-current density.
The choice $g=\epsilon_{0}c/\hbar$ in (\ref{Maxwellcontinuity}) is also
consistent with the QFT expression (\ref{QEDproduct}) that describes the
orthogonality of the localized photon states at $\left(  t,\mathbf{x}\right)
$ and $\left(  t,\mathbf{y}\right)  $.

With $g=\epsilon_{0}c/\hbar$ and $\widetilde{A}=A_{c}$ the photon four-current
(\ref{Maxwellcontinuity}) becomes%
\begin{equation}
J^{\mu}\left(  x\right)  =-\frac{\mathrm{i}\epsilon_{0}c}{\hbar}A_{\nu}^{\ast
}\left(  x\right)  \overleftrightarrow{\partial}^{\mu}A_{c}^{\nu}\left(
x\right)  . \label{Jph}%
\end{equation}
where $A$ and $A_{c}$ satisfy (\ref{MaxwellWaveEq}). {{The minus sign{ in
(\ref{Jph})} ensures that its space-like terms are positive while its
time-like term is negative, so that the scalar and longitudinal terms make no
net contribution to }}$J^{0}\left(  x\right)  ${{.}}

In the basis of positive and negative frequency definite helicity states,
(\ref{Afield}), substitution of (\ref{Ac}) in (\ref{Jph}) gives%
\begin{equation}
J^{0}\left(  x\right)  =\frac{2\epsilon_{0}c}{\hbar}\sum_{\epsilon,\lambda
=\pm}A_{\lambda\nu}^{\epsilon\ast}\left(  x\right)  \widehat{D}^{1/2}%
A_{\lambda}^{\epsilon\nu}\left(  x\right)  \label{DensityInEpsilonBasis}%
\end{equation}
where we have used the fact that the two terms in (\ref{Jph}) are equal if
$\epsilon^{\prime}=\epsilon$ and cancel if $\epsilon^{\prime}=-\epsilon$. From
(\ref{epsilon}) $\mathrm{i}\partial_{ct}A^{\epsilon}\equiv\epsilon
\widehat{D}^{-1/2}A^{\epsilon}$, so (\ref{DensityInEpsilonBasis}) is a sum of
positive and negative frequenc$\epsilon$y terms, in contrast to the standard
(charge) density that is their difference.

The{{\ photon inner product on a hyperplane of simultaneity at instant }}$t${{
will be defined as}}%
\begin{equation}
\left(  A_{1\lambda}^{\epsilon},A_{2\lambda^{\prime}}^{\epsilon^{\prime}%
}\right)  =-\frac{\mathrm{i}\epsilon_{0}c}{\hbar}\int_{t}\mathrm{d}%
\mathbf{x}A_{1\lambda\nu}^{\epsilon\ast}\left(  x\right)
\overleftrightarrow{\partial}_{ct}A_{c2\lambda^{\prime}}^{\epsilon^{\prime}%
\nu}\left(  x\right)  . \label{innerproductont}%
\end{equation}
It can be generalized to the covariant form%
\begin{equation}
\left(  A_{1\lambda}^{\epsilon},A_{2\lambda^{\prime}}^{\epsilon^{\prime}%
}\right)  _{n}=-\frac{\mathrm{i}\epsilon_{0}c}{\hbar}\int\mathrm{d}\sigma
n_{\mu}\,A_{1\lambda\nu}^{\epsilon\ast}\left(  x\right)
\overleftrightarrow{\partial}^{\mu}A_{c2\lambda^{\prime}}^{\epsilon^{\prime
}\nu}\left(  x\right)  \label{covariantinnerproduct}%
\end{equation}
on an arbitrary spacelike hyperplane with normal $n_{\mu}$
\cite{MostafazadehZamani,HawtonDebierre}. The choice $n=\left(
1,0,0,0\right)  $ gives (\ref{innerproductont}).

Substitution of (\ref{Ac}) and the invariant plane wave normalization
(\ref{korthogonal}) in (\ref{innerproductont}) with $c\widehat{D}%
^{1/2}\rightarrow\omega_{k}$ gives{{%
\begin{align}
\left(  A_{1\lambda}^{\epsilon},A_{2\lambda^{\prime}}^{\epsilon^{\prime}%
}\right)   &  =-\frac{2\epsilon_{0}c}{\hbar}\int_{t}\mathrm{d}\mathbf{x}%
A_{1\lambda\nu}^{\epsilon\ast}\left(  x\right)  \widehat{D}^{1/2}A_{2\lambda
}^{\epsilon\nu}\left(  x\right)  \delta_{\epsilon,\epsilon^{\prime}%
}\label{photonxspaceinnerproduct}\\
&  =\frac{2\epsilon_{0}c}{\hbar}\left\langle A_{1\lambda}^{\epsilon
}|\widehat{D}^{1/2}A_{2\lambda}^{\epsilon}\right\rangle \delta_{\epsilon
,\epsilon^{\prime}}\delta_{\lambda,\lambda^{\prime}}\zeta_{\lambda}.
\label{photoninnerproduct}%
\end{align}
The integrand of (}}\ref{photonxspaceinnerproduct}{{) is just photon density
in the form (\ref{DensityInEpsilonBasis}) if }}$A_{2\lambda^{\prime}%
}^{\epsilon^{\prime}}=A_{1\lambda}^{\epsilon}=A_{\lambda}^{\epsilon}$. {{In
the second line the sum over }}$\nu$ has been evaluated to give {t{he factor
}}$-\delta_{\lambda,\lambda^{\prime}}\zeta_{\lambda}$ that follows from
(\ref{es}).

The configuration space four-potential (\ref{Afield}) can be generalized to a
state vector in an arbitrary basis in a manner analogous to the vector
notation used in three dimensions \cite{CT}. Eq. (\ref{Afield}) is isomorphic
to the state vector%
\begin{equation}
\left\vert A_{\lambda}^{\epsilon\mu}\right\rangle =\mathrm{i}\sqrt{\frac
{\hbar}{\epsilon_{0}}}\int_{t}\frac{\mathrm{d}\mathbf{k}}{\left(  2\pi\right)
^{3}2\omega_{k}}c_{\lambda}^{\epsilon\mu}\left(  \mathbf{k}\right)  \left\vert
1_{\mathbf{k}\lambda}\right\rangle . \label{Aket}%
\end{equation}
Substitution of (\ref{Aket}) in (\ref{photoninnerproduct}) with $c\widehat{D}%
^{1/2}\rightarrow\omega_{k}${{ gives}}%
\begin{equation}
\left(  A_{1\lambda}^{\epsilon},A_{2\lambda^{\prime}}^{\epsilon^{\prime}%
}\right)  =\int_{t}\frac{\mathrm{d}\mathbf{k}}{\left(  2\pi\right)  ^{3}%
}c_{1\lambda}^{\epsilon\ast}\left(  \mathbf{k}\right)  c_{2\lambda}^{\epsilon
}\left(  \mathbf{k}\right)  \delta_{\epsilon,\epsilon^{\prime}}\delta
_{\lambda,\lambda^{\prime}}\zeta_{\lambda}. \label{photonkspaceinnerproduct}%
\end{equation}
With the more common non-invariant plane wave normalization denoted by the
subscript NW (Newton-Wigner),%
\begin{equation}
\left\langle 1_{\mathbf{k}\lambda}|1_{\mathbf{q}\lambda^{\prime}}\right\rangle
_{NW}=\delta_{\lambda,\lambda^{\prime}}\left(  2\pi\right)  ^{3}\delta\left(
\mathbf{k}-\mathbf{q}\right)  , \label{NWknormalization}%
\end{equation}
the $\mathbf{k}$-space inner product that replaces
(\ref{photonkspaceinnerproduct}) is%
\begin{equation}
\left(  A_{1\lambda}^{\epsilon},A_{2\lambda^{\prime}}^{\epsilon^{\prime}%
}\right)  _{NW}=\int_{t}\frac{\mathrm{d}\mathbf{k}}{\left(  2\pi\right)
^{3}2\omega_{k}}c_{1\lambda}^{\epsilon\ast}\left(  \mathbf{k}\right)
c_{2\lambda}^{\epsilon}\left(  \mathbf{k}\right)  \delta_{\epsilon
,\epsilon^{\prime}}\delta_{\lambda,\lambda^{\prime}}\zeta_{\lambda}.
\label{NWinnerproduct}%
\end{equation}

{{All modes of the four-potential can be treated similarly except that }%
}$A^{0}$ makes a negative contribution to the inner product, while its spatial
components make positive contributions. Since (\ref{Lorenz}) and
(\ref{1stQuantizedLorenz}) imply equal numbers of scalar and longitudinal
photons, their contributions cancel.{ States }$\left\vert n\right\rangle ${
containing no transverse modes but only }$n$ scalar/longitudual pairs are
equivalent to the vacuum so their contribution to the Fock space inner product
is $\left\langle n|n\right\rangle =\left\langle 0|0\right\rangle $
\cite{ItzyksonZuber}. The one-photon inner product $\left(  A_{1\lambda
}^{\epsilon},A_{2\lambda^{\prime}}^{\epsilon^{\prime}}\right)  $ does not
count these scalar/longitudual pairs and contains no vacuum contribution so
only transverse photons are counted.

For transverse modes the boundary condition at $t_{0}$ can be written as
\begin{align}
\mathbf{A}_{\lambda}\left(  t_{0},\mathbf{x}\right)   &  =\mathbf{A}%
_{\lambda0}\left(  \mathbf{x}\right)  ,\label{Ainitial}\\
\mathbf{E}_{\lambda}\left(  t_{0},\mathbf{x}\right)   &  =\mathbf{E}%
_{\lambda0}\left(  \mathbf{x}\right)  , \label{Einitial}%
\end{align}
where $\mathbf{E}_{\lambda}\left(  t,\mathbf{x}\right)  =-\partial
_{t}\mathbf{A}_{\lambda}\left(  t,\mathbf{x}\right)  $ for $\lambda=1,2$. This
is analogous to imposing initial conditions on the displacement and velocity
of an infinitesiment element of a vibrating string or membrane.

\subsection{Photon observables}

In this subsection we will find the eigenvectors and eigenvalues of the
momentum, energy, position, and angular momentum operators and write the
position space probability amplitude as the projection of an arbitrary state
vector onto bases of position eigenvectors and obtain real functions by taking
the real parts of the $\epsilon=+$ fields and probability amplitude. Examples
of circular and linear polarization, plane waves, and propagation toward and
away from a point of localization are given.

In $\mathbf{k}$-space the momentum operator is $\widehat{\mathbf{P}}%
=\hbar\mathbf{k}$ and the photon position operator with commuting components,
$\widehat{\mathbf{x}}$, is related to the position operator $\mathrm{i}%
\mathbf{\partial}_{\mathbf{k}}$ as $\widehat{\mathbf{x}}=\widehat{R}k^{\alpha
}\mathrm{i}\mathbf{\partial}_{\mathbf{k}}k^{-\alpha}\widehat{R}^{-1}$ where
\cite{HawtonBaylis}
\begin{equation}
\widehat{R}=\exp\left(  -\mathrm{i}\widehat{\sigma}\chi\right)  \exp\left(
-\mathrm{i}\widehat{S}_{3}\phi\right)  \exp\left(  -\mathrm{i}\widehat{S}%
_{2}\theta\right)  . \label{R}%
\end{equation}
The operator $\widehat{R}$ rotates $\mathbf{e}_{1}+\mathrm{i}\sigma
\mathbf{e}_{2}$ about $\mathbf{e}_{2}$ by $\theta$, then about $\mathbf{e}%
_{3}$ by $\phi$ to give $\mathbf{e}_{\theta}+\mathrm{i}\sigma\mathbf{e}_{\phi
}$, and finally about $\mathbf{e}_{\mathbf{k}}$ by $\chi$. Here $\partial
_{\mathbf{k}}$ is the $\mathbf{k}$-space gradient, $\widehat{S}_{i}$ are the
Cartesian components of the spin operator $\widehat{\mathbf{S}}$,
\begin{equation}
\widehat{\sigma}=\mathbf{e}_{\mathbf{k}}\cdot\widehat{\mathbf{S}},
\label{helicity}%
\end{equation}
is the helicity operator, $\theta$ and $\phi$ are the $\mathbf{k}$-space
spherical polar angles, $\chi\left(  \theta,\phi\right)  $ is the Euler angle
and the $\mathbf{k}$-space spherical polar unit vectors are $\mathbf{e}%
_{\theta},$ $\mathbf{e}_{\phi}$ and $\mathbf{e}_{\mathbf{k}}$. The transverse
unit vectors $\widehat{R}\left(  \mathbf{e}_{1}+\mathrm{i}\lambda
\mathbf{e}_{2}\right)  /\sqrt{2}$ with helicity $\lambda=\pm1$ are
\begin{equation}
\mathbf{e}_{\lambda}^{\left(  \chi\right)  }=\frac{1}{\sqrt{2}}\left(
\mathbf{e}_{\theta}+\mathrm{i}\lambda\mathbf{e}_{\phi}\right)  \mathrm{e}%
^{-\mathrm{i}\lambda\chi}. \label{transverseeigenvectors}%
\end{equation}
The longitudinal unit vector is $R\mathbf{e}_{3}=\mathbf{e}_{\mathbf{k}}$ and
the scalar unit vector that is unrotated by $R$ is given by (\ref{e0}).

The photon position operator with commuting components derived in
\cite{HawtonPosOp} is
\begin{equation}
\widehat{\mathbf{x}}^{\left(  \alpha\right)  }=\mathrm{i}\mathbf{\partial
}_{\mathbf{k}}-\mathrm{i}\alpha\frac{\mathbf{k}}{\left\vert \mathbf{k}%
\right\vert ^{2}}+\frac{1}{\left\vert \mathbf{k}\right\vert ^{2}%
}\mathbf{k\times}\widehat{\mathbf{S}}-\widehat{\sigma}\mathbf{a}\left(
\theta,\phi\right)  . \label{x}%
\end{equation}
The parameter $\alpha$ was introduced for convenience so that both
Newton-Wigner $\left(  \alpha=1/2\right)  $ and covariant $\left(
\alpha=0\right)  $ position eigenvectors could be included in a single
expression. If $\alpha=1/2$, (\ref{x}) equals the Pryce position operator plus
a term proportional to%
\begin{equation}
\mathbf{a}=\frac{\cos\theta}{k\sin\theta}\mathbf{e}_{\phi}+\mathbf{\partial
}_{\mathbf{k}}\chi. \label{a}%
\end{equation}
It is this additional term that gives $\widehat{\mathbf{x}}^{\left(
\alpha\right)  }$ commuting components. The Euler angle $\chi\left(
\theta,\phi\right)  $ is defined as a general rotation about $\mathbf{k}$. Any
possible transverse basis is the set of eigenvectors of (\ref{x}) for some
$\chi\left(  \theta,\phi\right)  $. Since experiments are often performed on
optical beams with definite angular momentum, the case $\chi=-m\phi$ for which
the position eigenvectors have intrinsic angular momentum $\hbar m\sigma$ in
some arbitrary but fixed direction is of special interest. For this choice of
$\chi$ (\ref{a}) becomes
\begin{equation}
\mathbf{a}^{\left(  m\right)  }=\frac{\cos\theta-m}{k\sin\theta}%
\mathbf{e}_{\phi}. \label{am}%
\end{equation}

An alternative to specifying the $t$-hyperplane is to work in the
Schr\"{o}dinger picture where operators and their eigenvectors are time
independent and the time dependence is contained in the state vector
\cite{SakuraiModernQM}. Transformation between the Schr\"{o}dinger and
Heisenberg pictures can be performed using a unitary time evolution operator.
In the $\epsilon=\pm$ basis time evolution is determined by \textrm{i}%
$\partial_{t}A_{\lambda}^{\epsilon}\left(  x\right)  =\epsilon c\widehat{D}%
^{1/2}A_{\lambda}^{\epsilon}\left(  x\right)  $ that can be integrated from
$t$ to $t+\tau$ to give%
\begin{align}
A_{\lambda}^{\epsilon}\left(  t+\tau,\mathbf{x}\right)   &  =\widehat{U}%
^{\epsilon}\left(  \tau\right)  A_{\lambda}^{\epsilon}\left(  t,\mathbf{x}%
\right)  ,\label{timeevolution}\\
\widehat{U}^{\epsilon}\left(  \tau\right)   &  =e^{-\mathrm{i}\epsilon
c\widehat{D}^{1/2}\tau}, \label{U}%
\end{align}
where $\widehat{U}^{\epsilon}$ is a unitary operator satisfying $\widehat{U}%
^{\dagger}=\widehat{U}^{-1}$ so that $\widehat{U}\widehat{U}^{\dagger
}=\widehat{U}^{\dagger}\widehat{U}=\widehat{1}$.

The operator equation for a state vector with momentum $\hbar\mathbf{q}$ and
helicity $\lambda$ is
\begin{equation}
\widehat{\mathbf{P}}\left\vert 1_{\lambda\mathbf{q}}\right\rangle
=\hbar\mathbf{q}\left\vert 1_{\lambda\mathbf{q}}\right\rangle .
\label{momentumeigenvectors}%
\end{equation}
These states will be normalized according to (\ref{korthogonal}). The basis of
momentum eigenvectors is orthogonal, Dirac $\delta$-function normalized and
complete \cite{CT}. Thus any state $\left\vert A\right\rangle $ can be
described by an integral over these plane waves. The Hamiltonian operator%
\begin{equation}
\widehat{H}=\hbar c\widehat{D}^{1/2} \label{H}%
\end{equation}
has positive energy eigenvectors, $\hbar c\left\vert \mathbf{k}\right\vert
=\hbar\omega_{k}$.

The position space probability amplitude is the projection of a particle's
state vector onto a basis of position eigenvectors. In the Schr\"{o}dinger
picture the $\mathbf{k}$-space representation of the position operator is
(\ref{x}) and the eigenvector equation and position eigenvector at
$\mathbf{y}$ are
\begin{align}
\widehat{\mathbf{x}}^{\left(  \alpha\right)  }c_{\lambda\mathbf{y}}^{\left(
\alpha\right)  }  &  =\mathbf{y}c_{\lambda\mathbf{y}}^{\left(  \alpha\right)
},\label{xeveceq}\\
c_{\lambda\mathbf{y}}^{\left(  \alpha\right)  \mu}\left(  \mathbf{k}\right)
&  =e_{\lambda}^{\mu\ast}\left(  \mathbf{k}\right)  \omega_{k}^{\alpha
}e^{-\mathrm{i}\epsilon\mathbf{k}\cdot\mathbf{y}}. \label{evecs}%
\end{align}
The choice $\alpha=0$ corresponds to the invariant normalization
(\ref{korthogonal}) with inner product (\ref{photonkspaceinnerproduct}), while
the Newton-Wigner choice $\alpha=1/2$ requires (\ref{NWknormalization}) and
(\ref{NWinnerproduct}). In either case, for $\lambda=\pm1$,
\begin{equation}
\left(  A_{\mathbf{x}\lambda}^{\epsilon},A_{\mathbf{y}\lambda^{\prime}%
}^{\epsilon^{\prime}}\right)  =\delta_{\lambda,\lambda^{\prime}}%
\delta_{\epsilon,\epsilon^{\prime}}\delta\left(  \mathbf{x}-\mathbf{y}\right)
. \label{xeigenvectororthogonality}%
\end{equation}
The factor $\delta\left(  \mathbf{x}-\mathbf{y}\right)  $ is infinite at
$\mathbf{x}=\mathbf{y}$, that is the position eigenvectors, like the momentum
eigenvectors, are not normalizable. The form of
(\ref{photonkspaceinnerproduct}) is due to the invariant plane wave
normalization (\ref{korthogonal}) that introduces a factor $\omega_{k}$ into
the inner product (\ref{photoninnerproduct}). Since the covariant case
$\alpha=0$ is the focus of this paper, we will define $c_{\mathbf{y}\lambda
}=c_{\mathbf{y}\lambda}^{\left(  0\right)  }$. In this case the Heisenberg
picture $\mathbf{k}$-space position eigenvectors are the manifestly covariant
Lorentz four-vectors%
\begin{equation}
c_{x\lambda}^{\epsilon\mu}\left(  \mathbf{k}\right)  =\left\langle
1_{\mathbf{k}\lambda}|1_{x\lambda}^{\epsilon\mu}\right\rangle =e_{\lambda
}^{\mu\ast}\left(  \mathbf{k}\right)  e^{\mathrm{i}\epsilon\left(  \omega
_{k}t-\mathbf{k}\cdot\mathbf{x}\right)  }. \label{HPevecs}%
\end{equation}
Transformation to configuration space using the Lorentz invariant measure
\textrm{d}$\mathbf{k}/k$ yields a four-vector proportional to the
electromagnetic four-potential, (\ref{Afield}). The inverse Fourier transform
obtained using the trivial measure is the time derivative of the vector
potential, proportional to the electric field describing this instantaneously
localized position eigenvector. For definite helicity transverse modes
$\lambda=\pm$, $\mathbf{B=-}\mathrm{i}\lambda\mathbf{E}$, so this is the
Weber$\mathbf{\ }$vector, $\mathbf{E+}\mathrm{i}\lambda c\mathbf{B}%
=2\mathbf{E}$.

The eigenvalues are the possible observed values, so they must be real. Since
$c_{\mathbf{x}\lambda}^{\left(  \alpha\right)  }$ and $c_{\mathbf{y\lambda}%
}^{\left(  \alpha\right)  }$ are eigenvectors of $\widehat{\mathbf{x}%
}^{\left(  \alpha\right)  }$ and $\widehat{D}^{-1/2}$ is Hermitian, the
expectation value of the position operator gives%
\begin{align}
\left(  A_{\mathbf{x\lambda}}^{\epsilon},\widehat{\mathbf{x}}^{\left(
\alpha\right)  }A_{\mathbf{y\lambda}}^{\epsilon}\right)   &  =\mathbf{y}%
\left(  A_{\mathbf{x\lambda}}^{\epsilon},A_{\mathbf{y\lambda}}^{\epsilon
}\right)  ,\label{xopexpectation}\\
\left(  A_{\mathbf{x\lambda}}^{\epsilon},\widehat{\mathbf{x}}^{\left(
\alpha\right)  }A_{\mathbf{y\lambda}}^{\epsilon}\right)  ^{\ast}  &  =\left(
\widehat{\mathbf{x}}^{\left(  \alpha\right)  \dagger}A_{\mathbf{x\lambda}%
}^{\epsilon},A_{\mathbf{y\lambda}}^{\epsilon}\right)  ^{\ast}.\nonumber
\end{align}
If $\alpha=1/2,$ $\widehat{\mathbf{x}}^{\left(  1/2\right)  \dagger
}=\widehat{\mathbf{x}}^{\left(  1-1/2\right)  }=\widehat{\mathbf{x}}^{\left(
1/2\right)  }$ is the well known Newton-Wigner result. If $\alpha=0$ the inner
product is (\ref{photonkspaceinnerproduct}) and $\widehat{\mathbf{x}}^{\left(
0\right)  \dagger}=\widehat{\mathbf{x}}^{\left(  0\right)  }\equiv
\widehat{\mathbf{x}}$. In either case $\widehat{\mathbf{x}}^{\left(
\alpha\right)  }$ is Hermitian. Since according to
(\ref{xeigenvectororthogonality}) $\left(  A_{\mathbf{x\lambda}}^{\epsilon
},A_{\mathbf{y\lambda}}^{\epsilon}\right)  =\delta\left(  \mathbf{x}%
-\mathbf{y}\right)  $, it follows that
\begin{equation}
\left(  \mathbf{y}-\mathbf{x}^{\ast}\right)  \delta\left(  \mathbf{x}%
-\mathbf{y}\right)  =0, \label{real}%
\end{equation}
that is the eigenvalues of $\widehat{\mathbf{x}}^{\left(  \alpha\right)  }$
are real as expected for Hermitian operators.

The state vector (\ref{Aket}) can be projected onto the $\mathbf{k}$- space
basis using (\ref{korthogonal}). With $k=\left(  \omega_{k}/c,\mathbf{k}%
\right)  $ this gives%
\begin{equation}
A_{\lambda}^{\epsilon}\left(  k\right)  =\mathrm{i}\sqrt{\hbar}c_{\lambda
}^{\epsilon}\left(  \mathbf{k}\right)  . \label{KGkspacepotential}%
\end{equation}
The velocity operator%
\begin{equation}
\widehat{\mathbf{v}}=\overset{.}{\widehat{\mathbf{x}}}=\frac{1}{\mathrm{i}%
\hbar}\left[  \widehat{\mathbf{v}},\widehat{H}\right]  \label{v}%
\end{equation}
equals $c\mathbf{e}_{k}$ in $\mathbf{k}$-space.

The transverse photon position eigenvectors have a definite three\textbf{-}%
component of total angular momentum with indefinite spin and orbital
contributions. Writing the total angular momentum as the sum of its extrinsic
and intrinsic parts,
\begin{align}
\widehat{\mathbf{J}}  &  =\hbar\widehat{\mathbf{x}}\times\widehat{\mathbf{k}%
}+\widehat{\mathbf{J}}_{int},\label{J}\\
\widehat{\mathbf{J}}_{int}  &  =\hbar\lambda\left(  \frac{\cos\theta-m}%
{\sin\theta}\mathbf{e}_{\theta}\mathbf{+e}_{\mathbf{k}}\right)  . \label{Jint}%
\end{align}
Using $\mathbf{e}_{\theta}\cdot\mathbf{e}_{3}=-\sin\theta$ and $\mathbf{e}%
_{\mathbf{k}}\cdot\mathbf{e}_{3}=\cos\theta$,
\begin{equation}
J_{int,3}=\hbar m\lambda. \label{J3}%
\end{equation}
In Cartesian components%
\begin{align}
\mathbf{e}_{\lambda}  &  =\frac{1}{2\sqrt{2}}\left[  \left(  \cos
\theta-\lambda\right)  e^{\mathrm{i}\left(  m\lambda+1\right)  \phi}\left(
\mathbf{e}_{1}-\mathrm{i}\lambda\mathbf{e}_{2}\right)  \right.  \label{ebasis}%
\\
&  -\left.  \sqrt{2}\sin\theta e^{\mathrm{i}m\lambda\phi}\mathbf{e}%
_{3}+\left(  \cos\theta+\lambda\right)  e^{\mathrm{i}\left(  m\lambda
-1\right)  \phi}\left(  \mathbf{e}_{1}+\mathrm{i}\lambda\mathbf{e}_{2}\right)
\right]  .\nonumber
\end{align}
With $J_{3}=L_{3}+S_{3}$ where $L_{3}$ and $S_{3}$ are spin and intrinsic
orbital angular momentum parallel to the $3$-axis, in the first term of
(\ref{ebasis}) $L_{3}=\hbar\left(  m\lambda+1\right)  $ and $S_{3}%
=-\hbar\lambda,$ in the second $L_{3}=\hbar m\lambda$ and $S_{3}=0,$ while in
the third term $L_{3}=\hbar\left(  m\lambda-1\right)  $ and $S_{3}%
=\hbar\lambda.$ In all terms their sum is $\hbar m\lambda$.

The position eigenvectors also form a basis where
(\ref{xeigenvectororthogonality}) is their closure relation \cite{CT}. The
four-potential and field obtained obtained by projection of (\ref{Aket}) onto
the basis of position eigenvectors $\left\vert 1_{\lambda x}^{\epsilon\mu
}\right\rangle $ is (\ref{Afield}). Since for transverse modes $\mathbf{E}%
_{\lambda}\left(  x\right)  =-\partial_{t}\mathbf{A}_{\lambda}\left(
x\right)  $,%
\begin{align}
\mathbf{A}_{\lambda}^{\epsilon}\left(  t,\mathbf{x}\right)   &  =\mathrm{i}%
\sqrt{\frac{\hbar}{\epsilon_{0}}}\int\frac{\mathrm{d}\mathbf{k\ }}{\left(
2\pi\right)  ^{3}2\omega_{k}}c_{\lambda}^{\epsilon}\left(  \mathbf{k}\right)
\mathbf{e}_{\lambda}\left(  \mathbf{k}\right)  e^{-{i}\epsilon\left(
\omega_{k}t-\mathbf{k}\cdot\mathbf{x}\right)  },\label{A}\\
\mathbf{E}_{\lambda}^{\epsilon}\left(  t,\mathbf{x}\right)   &  =\epsilon
\sqrt{\frac{\hbar}{\epsilon_{0}}}\int\frac{\mathrm{d}\mathbf{k\ }}{\left(
2\pi\right)  ^{3}2}c_{\lambda}^{\epsilon}\left(  \mathbf{k}\right)
\mathbf{e}_{\lambda}\left(  \mathbf{k}\right)  e^{-{i}\epsilon\left(
\omega_{k}t-\mathbf{k}\cdot\mathbf{x}\right)  }, \label{E}%
\end{align}
for $\lambda=1,2$. The position space photon probability amplitude $\left(
1_{\lambda x}^{\epsilon\mu},A\right)  $ is
\begin{equation}
\psi_{\lambda}^{\epsilon}\left(  x\right)  =\int\frac{\mathrm{d}\mathbf{k}%
}{\left(  2\pi\right)  ^{3}}e^{-\mathrm{i}\epsilon\left(  \omega
_{k}t-\mathbf{k}\cdot\mathbf{x}\right)  }c_{\lambda}^{\epsilon}\left(
\mathbf{k}\right)  . \label{Psi}%
\end{equation}

According to the Hegerfeldt theorem positive frequency fields\ do not
propagate causally. Eq. (\ref{deltaPV}) in Section II.B is an example that
illustrates this. As in classical electromagnetic theory it has been
convenient up to this point to use $\epsilon=\pm$ basis, but we will now
define the real fields and wave functions that propagate causally. For real
fields $A_{\lambda}^{-\mu}\left(  x\right)  =A_{\lambda}^{+\mu\ast}\left(
x\right)  $ so $A_{\lambda}^{-\mu}\left(  x\right)  $ is redundant and we will
define%
\begin{align}
A^{\mu}\left(  x\right)   &  =\operatorname{Re}\sum_{\lambda}A_{\lambda}%
^{+\mu}\left(  x\right)  ,\label{photonrealpotential}\\
\pi^{\mu}\left(  x\right)   &  =\operatorname{Re}\sum_{\lambda}\pi_{\lambda
}^{+\mu}\left(  x\right)  ,\label{photonrealfield}\\
\psi\left(  x\right)   &  =\operatorname{Re}\sum_{\lambda}\psi_{\lambda}%
^{+}\left(  x\right)  , \label{realphotonwavefunction}%
\end{align}
where $c_{\lambda}^{-}\left(  \mathbf{k}\right)  =0$ and $x=\left(
t,\mathbf{x}\right)  $.

Since $\mathbf{A}\left(  x\right)  $ is the indefinite time integral of
$-\mathbf{E}\left(  x\right)  $, we will consider only a few examples of
transverse $\mathbf{E}\left(  x\right)  $ and $\psi\left(  x\right)  $. Taking
$c_{\lambda}^{+}\left(  \mathbf{k}\right)  =c_{\lambda_{0}}\left(
\mathbf{k}\right)  \delta_{\lambda,\lambda_{0}}$ where $c_{\lambda_{0}}\left(
\mathbf{k}\right)  $ is assumed to be real for simplicity,
\begin{align}
\mathbf{E}_{\lambda_{0}}\left(  x\right)   &  =\sqrt{\frac{\hbar}{\epsilon
_{0}}}\operatorname{Re}\int\frac{\mathrm{d}\mathbf{k}}{\left(  2\pi\right)
^{3}}e^{-\mathrm{i}kx}c\left(  \mathbf{k}\right)  e_{\lambda_{0}}^{\mu}\left(
\mathbf{k}\right) \nonumber\\
&  =\sqrt{\frac{\hbar}{\epsilon_{0}}}\int\frac{\mathrm{d}\mathbf{k}}{\left(
2\pi\right)  ^{3}}\left[  e_{\theta}\left(  \mathbf{k}\right)  \cos\left(
kx\right)  \right. \nonumber\\
&  \left.  +\lambda_{0}e_{\phi}\left(  \mathbf{k}\right)  \sin\left(
kx\right)  \right]  c_{\lambda_{0}}\left(  \mathbf{k}\right)  , \label{CP}%
\end{align}
The field (\ref{CP}) is circularly polarized and rotates clockwise about
$\mathbf{k}$ if $\lambda=+1$ and counterclockwise if $\lambda=-1$. If
$c_{\lambda_{0}}\left(  \mathbf{k}\right)  =c_{\lambda_{0}}\delta\left(
\mathbf{k}-\mathbf{q}\right)  $ these are the circularly polarized plane
waves,%
\begin{align}
\mathbf{E}_{\lambda_{0}}\left(  x\right)   &  =\sqrt{\frac{\hbar}{\epsilon
_{0}}}c_{\lambda_{0}}\left[  \mathbf{e}_{\theta}\left(  \mathbf{k}\right)
\cos\left(  kx\right)  +\lambda_{0}\mathbf{e}_{\phi}\left(  \mathbf{k}\right)
\sin\left(  kx\right)  \right]  ,\label{Rotating}\\
\psi_{\lambda_{0}}\left(  x\right)   &  =c_{\lambda_{0}}\left(  \mathbf{k}%
\right)  \left[  \cos\left(  kx\right)  +\lambda_{0}\sin\left(  kx\right)
\right]  . \label{PsiCP}%
\end{align}

Circularly polarized states can be added and subtracted to give linearly
polarization. If $c_{\lambda}^{\epsilon}\left(  \mathbf{k}\right)  =c_{\theta
}\left(  \mathbf{k}\right)  \delta_{\epsilon,+}\left(  \delta_{\lambda
,+}+\delta_{\lambda,-}\right)  /\sqrt{2}$ where $c_{\theta}\left(
\mathbf{k}\right)  $ is real,%
\begin{align}
\mathbf{E}_{\theta}\left(  x\right)   &  =\sqrt{\frac{\hbar}{\epsilon_{0}}%
}\operatorname{Re}\int\frac{\mathrm{d}\mathbf{k}}{\left(  2\pi\right)  ^{3}%
}e^{-\mathrm{i}kx}c_{\theta}\left(  \mathbf{k}\right)  \frac{\mathbf{e}%
_{+}\left(  \mathbf{k}\right)  +\mathbf{e}_{+}\left(  \mathbf{k}\right)
}{\sqrt{2}}\nonumber\\
&  =\sqrt{\frac{\hbar}{\epsilon_{0}}}\int\frac{\mathrm{d}\mathbf{k}}{\left(
2\pi\right)  ^{3}}\mathbf{e}_{\theta}\left(  \mathbf{k}\right)  \cos\left(
kx\right)  c_{\theta}\left(  \mathbf{k}\right)  \label{LP}%
\end{align}
describes a linearly polarized photon. The choice $c_{\lambda}^{\epsilon
}\left(  \mathbf{k}\right)  =c\left(  \mathbf{k}\right)  \delta_{\epsilon
,+}\left(  \delta_{\lambda,+}-\delta_{\lambda,-}\right)  /\left(
\mathrm{i}\sqrt{2}\right)  $ gives%
\begin{equation}
\mathbf{E}_{\phi}\left(  x\right)  =\sqrt{\frac{\hbar}{\epsilon_{0}}}\int%
\frac{\mathrm{d}\mathbf{k}}{\left(  2\pi\right)  ^{3}}\mathbf{e}_{\phi}\left(
\mathbf{k}\right)  \cos\left(  kx\right)  c_{\phi}\left(  \mathbf{k}\right)  .
\label{LP2}%
\end{equation}
If $c_{\theta}\left(  \mathbf{k}\right)  =c_{1}\delta\left(  \mathbf{k}%
-q\mathbf{e}_{3}\right)  $ and $c_{\phi}\left(  \mathbf{k}\right)
=c_{2}\delta\left(  \mathbf{k}-q\mathbf{e}_{3}\right)  $, these are the
linearly polarized plane waves%
\begin{align}
\mathbf{E}_{1}\left(  x\right)   &  =c_{1}\mathbf{e}_{1}\left(  \mathbf{k}%
\right)  \cos\left(  kx\right)  ,\label{Linear1}\\
\mathbf{E}_{2}\left(  x\right)   &  =c_{2}e_{2}\left(  \mathbf{k}\right)
\cos\left(  kx\right)  . \label{Linear2}%
\end{align}
The plane wave states (\ref{Rotating}), (\ref{Linear1}) and (\ref{Linear2})
are not normalizable.

A photon localized at $\mathbf{y}$ on the $t_{0}=0$ hyperplane is described by
$c_{\mathbf{y}\lambda}\left(  \mathbf{k}\right)  =e^{-\mathrm{i}%
\mathbf{k}\cdot\mathbf{y}}$ so that $\psi_{\mathbf{y}\lambda}\left(  x\right)
=\delta\left(  \mathbf{x}-\mathbf{y}\right)  $ at $t=0$. An explicit
expression for its time evolution can be obtained by integrating (\ref{PsiCP})
as in (\ref{deltaPV}) and \cite{HawtonDebierre} to give
\begin{equation}
\psi_{\mathbf{y}\lambda}\left(  x\right)  =\partial_{ct}\left[  \frac
{\delta\left(  r-ct\right)  -\delta\left(  r+ct\right)  }{4\pi r}\right]  .
\label{InOut}%
\end{equation}
It is infinite at $r=\pm ct$ and not normalizable. Thus it does not allow
calculation of a photon probability density, but it can be seen by inspection
that it propagates causally, inward on a spherical shell if $t<0$ and outward
if $t>0$.

Just as in Schr\"{o}dinger quantum mechanics, the position and momentum
eigenvectors are not square integrable but they form a basis and hence satisfy
a completeness relation \cite{CT}. Using (\ref{photonkspaceinnerproduct}),
(\ref{realphotonwavefunction}) and the Parserval-Plancherel identity the
squared norm of the state vector $\left\vert A\right\rangle $ can be written
as
\begin{align}
\left(  A,A\right)   &  =\int\frac{\mathrm{d}\mathbf{k}}{\left(  2\pi\right)
^{3}}c\left(  \mathbf{k}\right)  ^{2}\label{Norm2}\\
&  =\int\mathrm{d}\mathbf{x}\psi\left(  x\right)  ^{2}. \label{PsiNorm2}%
\end{align}
If $\left(  A,A\right)  $ is finite, $\left\vert A\right\rangle $ is
normalizable and the $\mathbf{x}$-space probability density is
\begin{equation}
\rho\left(  x\right)  =\psi\left(  x\right)  ^{2}. \label{photonxdensity}%
\end{equation}
This is the Born rule. In a normalized one photon state, $\left(  A,A\right)
=1$. The plane wave basis satisfies an analogous completeness relation that
gives a $\mathbf{k}$-space density $\rho\left(  k\right)  $.

Babaei and Mostafazadeh defined a photon inner product
\cite{BabaeiMostafazadeh} that is closely related to (\ref{photoninnerproduct}%
). The parameter $g$, called $l$ in \cite{BabaeiMostafazadeh}, is unspecified
and they use a real symmetric/antisymmetric basis \ written in terms of
$\mathbf{A}$ and $\partial_{t}\mathbf{A}\propto\mathbf{E}$. Their position
eigenvectors are of the Newton-Wigner form, the Coulomb gauge is used, and
their photon number density is not the time-like component of a conserved
four-current. They impose initial conditions on $\mathbf{A}$ and $\mathbf{E}$
as in (\ref{Ainitial}) and (\ref{Einitial}) and derive a position operator
that is the Heisenberg picture equivalent of (\ref{x}).

\section{Conclusion}

In this paper we have derived a manifestly covariant quantum mechanical theory
of the photon. The solutions to their equations of motion are real but, as in
classical electromagnetic theory, we found it convenient to define the Hilbert
space as a basis of positive and negative frequency waves using the
Babaei-Mostafazadeh positive definite photon number density. These complex
solutions can then be used to calculate real fields and wave functions as a
final step. We extended the previous work by deriving a photon four-current in
the Lorenz gauge with real potentials, fields and probability amplitude. We
found for the first time that, with invariant plane wave normalization, the
position operator is Hermitian and orthogonality of its covariant eigenvectors
can be verified directly. The guiding principles of manifest covariance and
consistency with QFT led to use of the four-potential in the Lorenz gauge. The
photon inner product is then the sum over definite helicity transverse modes
where the contributions of scalar and longitudinal photons cancel. All
observables are described by Hermitian operators and the position probability
amplitude is the projection of the photon's state onto a basis of position eigenvectors.

We have avoided called $\psi\left(  x\right)  $ or any of the electromagnetic
fields "the wave function" since singling out any one of these entities seems
arbitrary. It was concluded here that all the mathematical apparatus and
physical interpretation of classical electromagnetic theory is inherited by
this first quantized theory of the photon. On top of this base is added a
conserved Lorentz four-current density, an invariant inner product, and
Hermitian operators representing observables.

According to the Reeh-Schlieder theorem \cite{RS} there are no local
annihilation or creation operators, so a photon cannot be destroyed locally.
In a source free region reality of the field and wave function implies no net
absorption. The real photon wave function $\psi_{y\lambda}\left(  x\right)
=\frac{1}{2}\left[  \left(  A_{x\lambda},A_{y}\right)  +\left(  A_{y\lambda
},A_{x}\right)  \right]  $ is an entangled sum of the probability amplitudes
for a photon emitted at $y$ and absorbed at $x$ or emitted at $x$ and absorbed
at $y$. This is of the form (\ref{deltaPV}) where $x$ and $y$ are on a
space-like hyperplane. Individually these terms are nonlocal,~only their real
sum propagates causally. While the Reeh-Schlieder theorem does not apply
directly to a first quantized theory that contains no particle creation or
annihilation operators, to provides important physical insight.

With the addition of photon probability density, Maxwell's theory based on
Faraday experiments \cite{FaradayMaxwell} can be thought of as the first
excursion into quantum mechanics. Maxwell's equations are the original field
theory, so it should perhaps not come as a surprise that they have a first
quantized interpretation. There is no quantum-classical divide since the
quantum theory is just an extension of the classical field theory. Since our
formalism is consistent with the quantum electrodynamic creation and
annihilation operators, it can be extended to coherent and Fock states. In the
absence of sources and sinks there is no net photon absorption since the field
is real and positive energy absorption is cancelled out by negative energy
emission. If the electric four-current density $J_{e}$ is nonzero, Maxwell's
wave equation is no longer source free and (\ref{MaxwellWaveEq}) is replaced
with $\square A^{\mu}\left(  x\right)  =-\mu_{0}J_{e}\left(  x\right)  $ so
nonzero net absorption is possible and the spatial integral of the inner
product (\ref{photonkspaceinnerproduct}) is not conserved.

\textit{Acknowledgement: }I thank Juan Leon for valuable discussion and for
convincing us to take Reeh-Schlieder seriously.

\end{document}